\documentclass[aps,pre,reprint,amsmath,amssymb,superscriptaddress,showpacs,floatfix]{revtex4-2}
\usepackage{float}
\usepackage{graphicx}
\usepackage{dcolumn}
\usepackage{bm}
\usepackage[colorlinks, allcolors=blue]{hyperref}
\usepackage[usenames, dvipsnames]{color}
\begin{document}
%%%%%%%%%%%%%%%%%%%%%%%%
%%%%%%%%%%%%%%%%%%%%%%%%%%%%%%%%%%%%%%%%%%%%%%%%%
%Title of paper
\title{Covariance Geometry of Basis-Resolved Low-Energy Wavefunctions in the Spin-$1/2$ Kitaev--Heisenberg Model}
\author{Sk Saniur Rahaman}
\email{sksrahaman@imsc.res.in}
\author{S. R. Hassan}
\email{shassan@imsc.res.in}
\affiliation{
The Institute of Mathematical Sciences,
A CI of Homi Bhabha National Institute,
Chennai 600113, India
}
\date{\today}
\begin{abstract}
We develop a basis-resolved covariance framework for investigating the
organization of low-energy many-body wavefunctions in the spin-$1/2$
Kitaev--Heisenberg model. By constructing covariance matrices from local-spin,
bond-correlation, and plaquette-flux representations of the low-energy
states, Principal Component Analysis (PCA) is employed to identify the
dominant collective covariance modes. We find a systematic evolution of the
covariance geometry across the phase diagram: magnetically ordered phases are
described by an essentially one-dimensional covariance manifold, conventional
magnetic phase boundaries exhibit competition between leading covariance
modes, whereas the Kitaev regimes develop intrinsically multidimensional
covariance geometry. Furthermore, the same many-body wavefunction produces
distinct covariance geometries in different operator representations,
demonstrating that covariance geometry is determined jointly by the quantum
state and the physical observables used to probe it. Shannon entropy and the
participation ratio provide quantitative measures of this evolution. The
present framework establishes basis-resolved covariance geometry as a
complementary statistical perspective for characterizing frustrated quantum
many-body systems beyond conventional order parameters.
\end{abstract}
\maketitle
%%%%%%%%%%%%%%%%%%%%%%%%%%%%%%%%%%%%%%%%%%%%%%%%%%%%%%%%%%%%%%%%%%%%%%%%%%
\section{Introduction}
%%%%%%%%%%%%%%%%%%%%%%%%%%%%%%%%%%%%%%%%%%%%%%%%%%%%%%%%%%%%%%%%%%%%%%%%%%

Understanding how complex quantum many-body wavefunctions organize their
microscopic quantum correlations remains one of the central challenges of
condensed matter physics
\cite{kohn1959analytic,banerjee2016proximate}. While the Landau theory of
spontaneous symmetry breaking successfully classifies conventional phases of
matter through local order parameters and broken symmetries
\cite{landau1936theory,landau1937theory}, a broad class of strongly
correlated quantum systems lies beyond this paradigm, where competing
interactions and strong quantum fluctuations suppress conventional order,
giving rise to quantum phases whose essential properties are encoded in the
structure of their many-body wavefunctions and correlations
\cite{anderson1973resonating,balents2010spin,savary2017quantum,takagi2019concept}.
Developing physically transparent approaches for characterizing such states
therefore remains an important problem in modern many-body physics
\cite{gotfryd2017phase,carleo2019netket,vicentini2022netket,wu2024variational}.

Among frustrated quantum magnets, the spin-$1/2$
Kitaev--Heisenberg model has emerged as one of the most important paradigms
for studying the competition between magnetic order and quantum
fluctuations
\cite{kitaev2006anyons,chaloupka2010kitaev,chaloupka2015hidden,rau2014generic,oitmaa2015phase,georgiou2024spin,hassan2013stable}.
By continuously interpolating between isotropic Heisenberg exchange and the
bond-dependent Kitaev interaction, the model hosts a rich phase diagram
consisting of N\'eel, ferromagnetic, zigzag, and stripy ordered phases
together with two quantum spin-liquid regions located near the pure Kitaev
limits
\cite{chaloupka2013zigzag,gotfryd2017phase,georgiou2024spin}. Because these
phases originate from competing interactions of different symmetry, the
Kitaev--Heisenberg model provides an ideal platform for investigating how the
microscopic organization of low-energy many-body wavefunctions evolves across
qualitatively different quantum phases.

The conventional magnetic phases are naturally characterized through
symmetry-breaking order parameters. Ferromagnetic and antiferromagnetic
states are identified by uniform and staggered magnetization, whereas zigzag
and stripy phases are distinguished by characteristic spatial spin patterns
\cite{hassan2013stable,nanda2020phases,janssen2019heisenberg,vasilchikova2022magnetic}.
The situation is fundamentally different near the Kitaev limits, where no
conventional symmetry is broken and the absence of magnetic order alone does
not establish the existence of a quantum spin liquid
\cite{balents2010spin,kitaev2006anyons}. Consequently, the Kitaev regime is
commonly investigated using more subtle probes, including spin correlations,
plaquette-flux operators, excitation spectra, dynamical structure factors,
and entanglement-related quantities
\cite{knolle2014dynamics,lahtinen2017short,iqbal2013gapless,gupta2026multiple}.
These complementary observables reveal different aspects of the underlying
quantum state that cannot be inferred from conventional order parameters
alone.

Recent years have witnessed the rapid application of machine-learning methods
to quantum many-body physics
\cite{carrasquilla2017machine,carrasquilla2020machine,melko2019restricted,carleo2019netket,vicentini2022netket}.
Among unsupervised learning techniques, Principal Component Analysis (PCA)
has attracted considerable attention because it provides an interpretable
decomposition of the covariance structure of high-dimensional datasets
without requiring prior knowledge of the phase diagram
\cite{costa2017principal,hu2017discovering,rahaman2023machine,haldar2024study,ho2026unsupervised}.
In many classical and quantum lattice models, the dominant principal
components reproduce conventional order parameters and successfully identify
symmetry-breaking phase transitions
\cite{costa2017principal,hu2017discovering,miyajima2023machine,okabe2026berezinskii}.
These successes demonstrate that PCA efficiently captures the dominant
statistical fluctuations of the underlying data.

However, the physical interpretation of PCA in frustrated quantum systems
remains much less clear. Principal Component Analysis does not directly probe
quantum entanglement, topological order, or fractionalized excitations.
Instead, it identifies the dominant covariance modes within a chosen
representation of the data
\cite{li2026learning,ho2026unsupervised}. Consequently, the physical
information extracted by PCA depends fundamentally on how the quantum states
are represented before the covariance matrix is constructed. Understanding
the relationship between the microscopic organization of many-body
wavefunctions, the choice of physical observables, and the resulting
covariance geometry is therefore essential for giving physical meaning to the
principal components, particularly in systems where conventional order
parameters are absent.

In contrast to previous PCA studies that analyze expectation values,
correlation functions, or classical configurations, we develop a
basis-resolved covariance framework in which the statistical ensemble is
constructed directly from the dominant computational-basis components of the
many-body wavefunctions. The ground state and first excited state are first
obtained using the Symmetry Adapted Lower Energy Subspace Diagonalisation
(SALE--SD) method. After expanding the wavefunctions in the computational
basis, the dominant basis configurations are retained through a truncation
procedure. Local-spin, bond-correlation, and plaquette-flux operators are
then projected onto these retained basis states to construct basis-resolved
feature vectors. The resulting covariance matrices therefore simultaneously
encode the spatial organization of the physical observables and the
probability distribution of the dominant many-body basis configurations.

The emphasis of the present work is therefore not on constructing new order
parameters, but on understanding how the statistical geometry of
basis-resolved many-body wavefunctions evolves across different quantum
phases. We demonstrate that the magnetically ordered phases are
characterized by an essentially one-dimensional covariance geometry
dominated by a single collective covariance mode, whereas the conventional
magnetic phase boundaries exhibit a redistribution of covariance variance
among competing modes while remaining effectively low dimensional. In
contrast, the Kitaev regimes develop intrinsically multidimensional
covariance geometry, reflecting the increased complexity of the low-energy
wavefunctions. Furthermore, we show that the same quantum wavefunction
generates distinct covariance geometries in the local-spin,
bond-correlation, and plaquette-flux representations, demonstrating that
covariance geometry is a property of both the quantum state and the operator
representation used to probe it. To quantify these observations, Shannon
entropy and the participation ratio derived from the covariance spectrum are
employed as compact measures of the complexity of the covariance geometry.

The remainder of the paper is organized as follows. Section~II introduces the
Kitaev--Heisenberg model, the basis-resolved feature construction, and the
covariance formalism. Section~III presents the microscopic organization of
the low-energy wavefunctions and investigates the evolution of the covariance
geometry across the phase diagram using local-spin, bond-correlation, and
plaquette-flux representations. Shannon entropy and the participation ratio
are then introduced as quantitative measures of the covariance geometry.
Finally, Section~IV summarizes the principal conclusions and discusses
possible extensions of the present framework to other strongly correlated
quantum many-body systems.

%%%%%%%%%%%%%%%%%%%%%%%%%%%%%%%%%%%%%%%%%%%%%%%%%%%%%%%%%%%%%%%%%%%%%%%%%%
\section{Model and Basis-Resolved Covariance Formalism}
\label{sec:model_and_method}
%%%%%%%%%%%%%%%%%%%%%%%%%%%%%%%%%%%%%%%%%%%%%%%%%%%%%%%%%%%%%%%%%%%%%%%%%%

%%%%%%%%%%%%%%%%%%%%%%%%%%%%%%%%%%%%%%%%%%%%%%%%%%%%%%%%%%%%%%%%%%%%%%%%%%
\subsection{Kitaev--Heisenberg Model}
%%%%%%%%%%%%%%%%%%%%%%%%%%%%%%%%%%%%%%%%%%%%%%%%%%%%%%%%%%%%%%%%%%%%%%%%%%
\begin{figure}[t]
\centering
\includegraphics[width=0.60\columnwidth]
{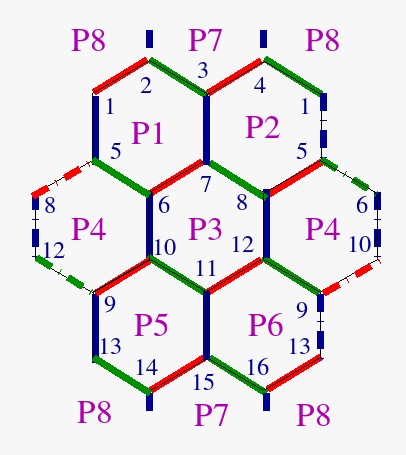}
\caption{(color online) 
Geometry of the periodic $N=16$ honeycomb cluster employed in the present
work. The red, green, and blue bonds denote the three inequivalent
bond-dependent next nearest neighbour Kitaev interactions $S_i^xS_j^x$,$S_i^yS_j^y$,
and $S_i^zS_j^z$,respectively. The elementary plaquettes,$P_1,\ldots,P_8$,
are labeled for constructing the plaquette-flux operators used in the
covariance analysis.}
\label{fig:schematic}
\end{figure}
The spin-$1/2$ Kitaev--Heisenberg model on the honeycomb lattice provides a
minimal framework for investigating the competition between conventional
isotropic magnetic exchange and bond-dependent anisotropic interactions.
Throughout this work we consider the periodic $N=16$ honeycomb cluster shown
in Fig.~\ref{fig:schematic}. The three nearest-neighbour bond directions are
distinguished according to the corresponding Kitaev interaction, while the
elementary plaquettes are labeled for constructing the plaquette-flux
operators introduced later.

The Hamiltonian of the spin-$1/2$ Kitaev--Heisenberg model is

\begin{equation}
\label{eqn:model_Kitaev_HB_HC}
H
=
J
\sum_{\langle ij\rangle}
{\bf S}_i\cdot{\bf S}_j
+
K
\sum_{\gamma=x,y,z}
\sum_{\langle ij\rangle_\gamma}
S_i^\gamma S_j^\gamma,
\end{equation}

where $\langle ij\rangle$ denotes nearest-neighbour lattice sites,
${\bf S}_i=(S_i^x,S_i^y,S_i^z)$ is the spin-$1/2$ operator at site $i$, and
$\langle ij\rangle_\gamma$ represents a nearest-neighbour bond of type
$\gamma=x,y,z$. The first term describes the isotropic Heisenberg exchange,
whereas the second term introduces the bond-dependent Kitaev interaction.

The exchange couplings are conveniently parameterized as
\begin{equation}
J=\cos\phi,
\qquad
K=\sin\phi,
\end{equation}
with $0\le \phi <2\pi$.

Varying the interaction parameter continuously interpolates between the
Heisenberg and Kitaev limits and generates a rich phase diagram consisting of
four magnetically ordered phases (N\'eel, zigzag, ferromagnetic, and stripy)
together with two Kitaev spin-liquid (KSL) regions near the pure Kitaev
limits.

The phase boundaries used throughout the present work are summarized in
Table~\ref{tab:phase_boundary}. These values are in good agreement with
previous numerical investigations of the Kitaev--Heisenberg model \cite{gotfryd2017phase,georgiou2024spin}. The
objective of the present work, however, is not to redetermine the phase
diagram itself, but to investigate how the organization of the low-energy
wavefunctions and their basis-resolved covariance geometry evolve across
these different regimes.

\begin{table}[t]
\centering
\caption{Phase boundaries of the spin-$1/2$ Kitaev--Heisenberg model used throughout the present covariance analysis.}
\label{tab:phase_boundary}
\begin{tabular}{lcc}
\hline\hline
Phase boundary & $\phi/\pi$ & $\phi$ ($\mathrm{deg}$) \\
\hline
N\'eel--KSL (AFM)      & 0.492 & 88.56 \\
KSL (AFM)--Zigzag      & 0.517 & 93.06 \\
Zigzag--FM             & 0.797 & 143.46 \\
FM--KSL (FM)           & 1.450 & 261.00 \\
KSL (FM)--Stripy       & 1.541 & 277.38 \\
Stripy--N\'eel         & 1.714 & 308.52 \\
\hline\hline
\end{tabular}
\end{table}

To investigate the evolution of the low-energy quantum states across these
phases, we first construct accurate low-energy eigenstates using the
Symmetry-Adapted Low-Energy Subspace Diagonalisation (SALE--SD) method,
which is briefly summarized below.

%%%%%%%%%%%%%%%%%%%%%%%%%%%%%%%%%%%%%%%%%%%%%%%%%%%%%%%%%%%%%%%%%%%%%%%%%%
\subsection{Symmetry-Adapted Low-Energy Subspace}
%%%%%%%%%%%%%%%%%%%%%%%%%%%%%%%%%%%%%%%%%%%%%%%%%%%%%%%%%%%%%%%%%%%%%%%%%%

The covariance analysis developed in this work requires an accurate
description of the low-energy eigenstates throughout the entire phase
diagram. Although exact diagonalization (ED) provides numerically exact
solutions, its computational cost grows exponentially with system size \cite{lanczos1950iteration,er1975iterativecalculationof}.
We therefore employ the Symmetry-Adapted Low-Energy Subspace
Diagonalisation (SALE--SD) method, which constructs a reduced Hilbert
space by exploiting the symmetries of the Hamiltonian together with an
iterative selection of energetically relevant basis states.

The resulting reduced Hilbert space preserves the essential low-energy
physics while significantly reducing the computational cost, thereby
enabling a systematic exploration of the complete phase diagram.

For each value of the interaction parameter $\phi$, the two lowest-energy
eigenstates are obtained by solving

\begin{equation}
H(\phi)
|\Psi_n(\phi)\rangle
=
E_n(\phi)
|\Psi_n(\phi)\rangle,
\qquad
n=0,1,
\label{eq:eigenvalue}
\end{equation}

where $|\Psi_0(\phi)\rangle$ and $|\Psi_1(\phi)\rangle$ denote the ground
state and first excited state, respectively.

\begin{figure}[t]
\centering
\includegraphics[width=0.99\columnwidth]
{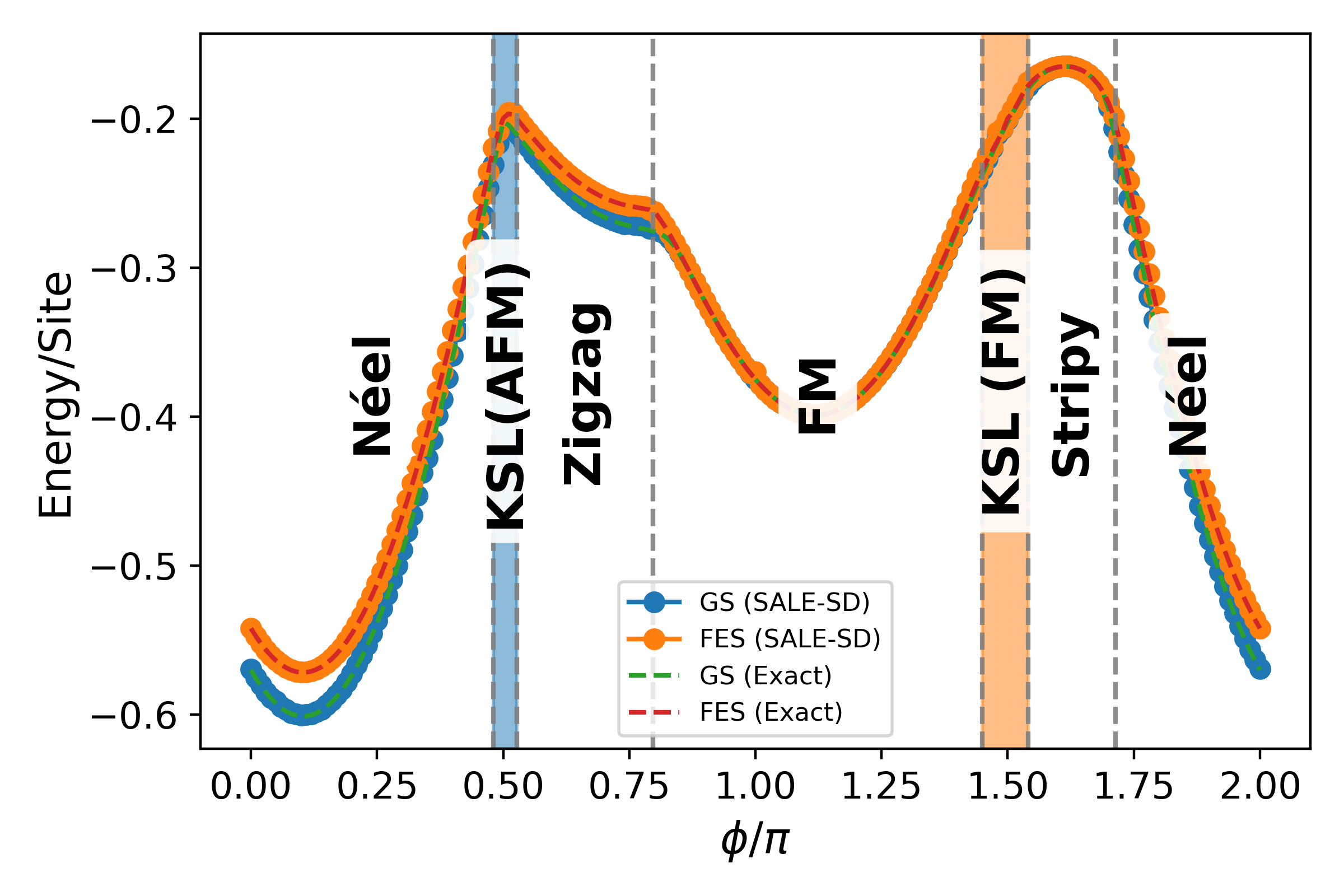}
\caption{Ground-state and first-excited-state energies per site for the
$N=16$ Kitaev--Heisenberg model obtained using the SALE--SD method (solid
lines) and exact diagonalization (dashed lines) as functions of the
interaction parameter $\phi$. The excellent agreement demonstrates that the
reduced symmetry-adapted basis accurately reproduces the low-energy spectrum
throughout the phase diagram.}
\label{fig:site_energy}
\end{figure}

Fig.\ref{fig:site_energy} compares the energies obtained using SALE--SD
with those from exact diagonalization over the full parameter range
$0\le\phi<2\pi$. The two calculations are essentially indistinguishable on
the scale of the figure for both the ground and first excited states,
demonstrating that the reduced symmetry-adapted basis faithfully reproduces
the low-energy spectrum throughout the phase diagram.

The collection of low-energy eigenstates
$\{|\Psi_n(\phi)\rangle\}$ generated in this manner forms the starting point
of the present investigation. Rather than analyzing individual eigenstates
through a small set of expectation values, we examine how their internal
structure evolves across the phase diagram and how this evolution is
reflected in different basis-resolved operator representations. To achieve
this, each eigenstate is first expressed in the computational basis, which
forms the foundation of the covariance formalism developed in the following
subsections.

%%%%%%%%%%%%%%%%%%%%%%%%%%%%%%%%%%%%%%%%%%%%%%%%%%%%%%%%%%%%%%%%%%%%%%%%%%
\subsection{Basis-Resolved Representation of the Wavefunction}
%%%%%%%%%%%%%%%%%%%%%%%%%%%%%%%%%%%%%%%%%%%%%%%%%%%%%%%%%%%%%%%%%%%%%%%%%%

The low-energy eigenstates obtained from SALE--SD constitute the starting
point of the present investigation. Rather than characterizing these states
through a limited set of expectation values, we begin from their complete
many-body wavefunctions expressed in the computational spin basis. For a
given interaction parameter $\phi$, the $n$th low-energy eigenstate is
expanded as

\begin{equation}
|\Psi_n(\phi)\rangle
=
\sum_{k}
C_k^{(n)}(\phi)
|\eta_k\rangle,
\label{eq:wavefunction}
\end{equation}

where
$\{|\eta_k\rangle\}$ denotes the complete computational basis and
$C_k^{(n)}(\phi)$ is the corresponding expansion coefficient.

The coefficients completely specify the quantum state. Their magnitudes
$|C_k^{(n)}|^2$ determine the statistical weight carried by individual basis
configurations, while their relative signs encode the quantum interference
between different many-body configurations. As the interaction parameter
$\phi$ is varied, both the amplitude distribution and the sign structure
evolve continuously, reflecting the reorganization of the underlying
low-energy wavefunction across the phase diagram.

In practice, the probability distribution
$|C_k^{(n)}|^2$
is highly inhomogeneous, with only a relatively small fraction of basis
configurations carrying significant weight. To construct an efficient yet
physically representative description of the wavefunction, the basis states
are ranked according to the magnitude of their coefficients, and the
dominant configurations are retained for subsequent analysis. Throughout
this work, we retain the $M=1000$ basis states with the largest coefficient
magnitudes for each low-energy eigenstate.

The retained basis is determined independently for every value of the
interaction parameter $\phi$. Consequently, the dominant basis
configurations evolve together with the wavefunction itself, allowing the
representation to adapt naturally to the changing character of the quantum
state across the phase diagram. The retained basis therefore represents the
dominant low-energy structure rather than a fixed set of computational basis
states.

The retained basis configurations constitute the statistical ensemble from
which the basis-resolved feature spaces are constructed. These feature
spaces provide physically meaningful representations of the same underlying
many-body wavefunction and form the input to the covariance analysis
developed below.

%%%%%%%%%%%%%%%%%%%%%%%%%%%%%%%%%%%%%%%%%%%%%%%%%%%%%%%%%%%%%%%%%%%%%%%%%%
\subsection{Local-Spin, Bond, and Plaquette-Flux Feature Spaces}
%%%%%%%%%%%%%%%%%%%%%%%%%%%%%%%%%%%%%%%%%%%%%%%%%%%%%%%%%%%%%%%%%%%%%%%%%%

To characterize the organization of the retained basis configurations, we
project the low-energy wavefunctions onto several physically relevant
operators. Each retained basis configuration therefore generates a feature
vector whose components are matrix elements of the chosen operator between
the computational basis and the many-body wavefunction.

Throughout this work we consider three complementary feature spaces:
local-spin, bond-correlation, and plaquette-flux representations.

For the local-spin representation, the feature associated with lattice site
$i$ is defined as

\begin{equation}
X_{ki}^{(S)}(\phi)
=C_k^*(\phi)
\langle
\eta_k
|
S_i^z
|
\Psi(\phi)
\rangle .
\label{eq:sitefeature}
\end{equation}
$C_k^*(\phi)$ is the complex conjugate of $C_k(\phi)$. Since the computational basis diagonalizes $S_i^z$, the matrix element
reduces to

\begin{equation}
X_{ki}^{(S)}(\phi)
=
|C_k(\phi)|^2\,
s_i^{(k)},
\end{equation}

where

\[
s_i^{(k)}
=
\pm\frac12
\]

is the spin projection of the retained basis configuration
$|\eta_k\rangle$ at lattice site $i$.

Thus, each retained basis configuration contributes its complete spin
pattern weighted by the corresponding wavefunction amplitude.

The bond-correlation representation is constructed from the nearest-neighbour
Heisenberg operator,

\begin{equation}
X_{k,ij}^{(B)}(\phi)
=C_k^*(\phi)
\langle
\eta_k
|
{\bf S}_i\cdot{\bf S}_j
|
\Psi(\phi)
\rangle .
\label{eq:bondfeature}
\end{equation}

Unlike the local-spin operator,
${\bf S}_i\cdot{\bf S}_j$
is not diagonal in the computational basis. Consequently,

\[
X_{k,ij}^{(B)}
=C_k^*(\phi)
\sum_l
C_l(\phi)
\,
\langle
\eta_k
|
{\bf S}_i\cdot{\bf S}_j
|
\eta_l
\rangle,
\]

so that the bond features incorporate both the wavefunction amplitudes and
the off-diagonal matrix elements of the bond operator.

To probe the bond-directional physics characteristic of the Kitaev model,
we further introduce a plaquette-flux representation based on the Wilson-loop
operator

\begin{equation}
\label{eqn:wilson_flux}
W_p
=
\sigma_1^x
\sigma_2^y
\sigma_3^z
\sigma_4^x
\sigma_5^y
\sigma_6^z,
\end{equation}

where the sequence of spin components follows the six bonds surrounding
plaquette $p$. The corresponding basis-resolved feature is defined as

\begin{equation}
X_{kp}^{(P)}(\phi)
=C_k^*(\phi)
\langle
\eta_k
|
W_p
|
\Psi(\phi)
\rangle .
\label{eq:plaqfeature}
\end{equation}

Since the Wilson-loop operator also contains off-diagonal spin operators,
its matrix elements generally involve contributions from multiple basis
configurations,

\[
X_{kp}^{(P)}
=C_k^*(\phi)
\sum_l
C_l(\phi)
\,
\langle
\eta_k
|
W_p
|
\eta_l
\rangle.
\]

The three feature spaces therefore probe complementary aspects of the same
underlying many-body wavefunction. The local-spin representation emphasizes
magnetic textures, the bond representation characterizes short-range spin
correlations, and the plaquette representation captures the local flux
degrees of freedom characteristic of the Kitaev limit. Together they provide
three distinct basis-resolved representations of the low-energy quantum
states that serve as the input for the covariance analysis.

%%%%%%%%%%%%%%%%%%%%%%%%%%%%%%%%%%%%%%%%%%%%%%%%%%%%%%%%%%%%%%%%%%%%%%%%%%
\subsection{Construction of the Covariance Matrix}
%%%%%%%%%%%%%%%%%%%%%%%%%%%%%%%%%%%%%%%%%%%%%%%%%%%%%%%%%%%%%%%%%%%%%%%%%%

The basis-resolved feature spaces introduced above provide a collection of
feature vectors representing the low-energy wavefunctions in different
operator representations. Each retained basis configuration contributes one
sample to the statistical ensemble.

To distinguish between the local basis index used in
Eq.~(\ref{eq:wavefunction}) and the complete statistical ensemble used for
the covariance analysis, we introduce a global sample index
$a=1,\ldots,N_{\mathrm{sam}}$, where each sample corresponds to one retained
basis configuration belonging to one of the low-energy eigenstates at a
particular value of the interaction parameter $\phi$. The total number of
samples is therefore denoted by $N_{\mathrm{sam}}$.

For every sample $a$, we construct a feature vector

\begin{equation}
{\bf X}_a
=
(X_{a1},X_{a2},\ldots,X_{ad}),
\end{equation}

where the feature dimension $d$ depends on the operator representation.
Specifically,

\[
d=
\begin{cases}
N, & \text{local-spin representation},\\
N_b, & \text{bond-correlation representation},\\
N_p, & \text{plaquette-flux representation},
\end{cases}
\]

with $N$, $N_b$, and $N_p$ denoting the numbers of lattice sites,
nearest-neighbour bonds, and plaquettes, respectively.

Collecting all samples produces the data matrix

\[
X=
\{X_{a\mu}\},
\qquad
a=1,\ldots,N_{\mathrm{sam}},
\quad
\mu=1,\ldots,d.
\]

Before constructing the covariance matrix, each feature is centered by
subtracting its ensemble average,

\begin{equation}
\widetilde X_{a\mu}
=
X_{a\mu}
-
\overline X_\mu,
\end{equation}

where

\begin{equation}
\overline X_\mu
=
\frac{1}{N_{\mathrm{sam}}}
\sum_{a=1}^{N_{\mathrm{sam}}}
X_{a\mu}.
\end{equation}

The covariance matrix is then defined as

\begin{equation}
C_{\mu\nu}
=
\frac{1}{N_{\mathrm{sam}}-1}
\sum_{a=1}^{N_{\mathrm{sam}}}
\widetilde X_{a\mu}
\widetilde X_{a\nu},
\label{eq:covariance}
\end{equation}

where the indices $\mu$ and $\nu$ refer to features within the chosen
operator representation.

Unlike the conventional quantum covariance matrix, which characterizes the
fluctuations of observables within a single quantum state, the covariance
matrix defined in Eq.~(\ref{eq:covariance}) describes the statistical
covariance of the basis-resolved feature ensemble constructed from the
low-energy wavefunctions sampled throughout the phase diagram. Consequently,
it quantifies how different physical features co-vary as the internal
structure of the low-energy wavefunctions evolves with the interaction
parameter.

For each operator representation, the covariance matrix therefore provides a
compact geometric description of the statistical organization of the
low-energy manifold. Strong off-diagonal matrix elements indicate collective
correlations between different physical features, whereas weak off-diagonal
elements correspond to nearly independent fluctuations. The dominant
collective structures contained in the covariance matrix are extracted
through its eigenvalue decomposition, which forms the basis of the principal
covariance analysis described in the following subsection.

%%%%%%%%%%%%%%%%%%%%%%%%%%%%%%%%%%%%%%%%%%%%%%%%%%%%%%%%%%%%%%%%%%%%%%%%%%
\subsection{Principal Covariance Modes and Quantified Projections}
%%%%%%%%%%%%%%%%%%%%%%%%%%%%%%%%%%%%%%%%%%%%%%%%%%%%%%%%%%%%%%%%%%%%%%%%%%

The dominant collective structures contained in the covariance matrix are
obtained by solving the eigenvalue problem

\begin{equation}
C{\bf w}_{\alpha}
=
\lambda_{\alpha}
{\bf w}_{\alpha},
\label{eq:pca}
\end{equation}

where $\lambda_{\alpha}$ is the eigenvalue associated with the
$\alpha$th covariance mode and
${\bf w}_{\alpha}$ is the corresponding normalized eigenvector.
Since the covariance matrix is real and symmetric, its eigenvectors form an
orthonormal basis of the chosen feature space. The eigenvalues measure the
variance captured by the corresponding covariance modes and therefore
quantify their relative importance in describing the statistical
organization of the basis-resolved ensemble.

Depending on the chosen feature representation, the covariance eigenvectors
describe collective patterns of local-spin configurations,
bond-correlation structures, or plaquette-flux distributions. The principal
covariance modes therefore provide the dominant directions along which the
low-energy wavefunctions exhibit the largest statistical variations.

To quantify the contribution of each sample to a given covariance mode, the
centered feature vector is projected onto the corresponding eigenvector,

\begin{equation}
P_{\alpha}(a)
=
{\bf w}_{\alpha}^{\,T}
\widetilde{\bf X}_{a},
\label{eq:projection}
\end{equation}

where $\widetilde{\bf X}_{a}$ denotes the centered feature vector of the
$a$th sample. The quantity
$P_{\alpha}(a)$
is the principal-component score of sample $a$ along the
$\alpha$th covariance mode.

Since many basis-resolved samples correspond to the same interaction
parameter $\phi$, it is convenient to characterize the overall strength of
each covariance mode by averaging the magnitudes of the corresponding
principal-component scores. We therefore define the quantified principal
component as
\begin{equation}
\label{eqn:Qalpha}
Q_\alpha(\phi) = 
\left\{
\begin{array}{l@{\quad}l}
\displaystyle\sum_{a\in\phi} |P_\alpha(a)|, & \text{local-spin}, \\
\displaystyle\sum_{a\in\phi} P_\alpha(a), & \text{bond-correlation}, \\
\displaystyle\sum_{a\in\phi} P_\alpha(a), & \text{plaquette-flux},
\end{array}
\right.
\end{equation}
where the summation extends over all retained basis configurations
associated with the selected low-energy eigenstates at interaction
parameter $\phi$, and $N_{\phi}$ denotes the corresponding number of
samples.

The quantity
$Q_{\alpha}(\phi)$
therefore measures the average contribution of the
$\alpha$th covariance mode to the basis-resolved ensemble at a given value
of the interaction parameter. Tracking
$Q_{\alpha}(\phi)$
throughout the phase diagram provides a compact description of how the
dominant collective covariance modes evolve as the low-energy wavefunctions
reorganize across different physical regimes.

Throughout the remainder of this work, three complementary pieces of
information are extracted from the covariance analysis: (i) the covariance
eigenvalue spectrum, which determines the number of statistically
significant collective modes; (ii) the covariance eigenvectors, which reveal
their physical structure in the chosen operator representation; and (iii)
the quantified principal components
$Q_{\alpha}(\phi)$,
which monitor the evolution of these modes across the phase diagram.
Together, these quantities provide a unified basis-resolved description of
the organization of the low-energy quantum states in local-spin,
bond-correlation, and plaquette-flux representations.

%%%%%%%%%%%%%%%%%%%%%%%%%%%%%%%%%%%%%%%%%%%%%%%%%%%%%%%%%%%%%%%%%%%%%%%%%%
\subsection{Computational Workflow}
%%%%%%%%%%%%%%%%%%%%%%%%%%%%%%%%%%%%%%%%%%%%%%%%%%%%%%%%%%%%%%%%%%%%%%%%%%

The complete basis-resolved covariance analysis developed in this work
consists of the following sequence of steps:

\begin{enumerate}

\item For each interaction parameter
$\phi$, the ground state and first excited state are computed using the
Symmetry-Adapted Low-Energy Subspace Diagonalisation (SALE--SD) method.

\item Each low-energy eigenstate is expanded in the computational spin basis,

\[
|\Psi_n(\phi)\rangle
=
\sum_k
C_k^{(n)}(\phi)
|\eta_k\rangle .
\]

\item The computational basis states are ranked according to
$|C_k^{(n)}|$, and the
$M=1000$
dominant basis configurations are retained.

\item For every retained basis configuration, basis-resolved feature vectors
are constructed in the chosen operator representation
(local-spin, bond-correlation, or plaquette-flux).

\item The feature vectors collected from all interaction parameters form the
statistical ensemble used to construct the centered data matrix.

\item The covariance matrix of the centered feature ensemble is computed and
diagonalized to obtain the principal covariance modes and their associated
eigenvalues.

\item Each centered feature vector is projected onto the covariance
eigenvectors to obtain the principal-component scores.

\item Finally, the quantified principal components
$Q_\alpha(\phi)$
are evaluated to monitor the evolution of the dominant covariance modes
throughout the phase diagram.

\end{enumerate}

This workflow is repeated independently for the local-spin,
bond-correlation, and plaquette-flux representations, allowing the same set
of low-energy wavefunctions to be analyzed from three complementary physical
perspectives.

%%%%%%%%%%%%%%%%%%%%%%%%%%%%%%%%%%%%%%%%%%%%%%%%%%%%%%%%%%%%%%%%%%%%%%%%%%
\section{Results and Discussion}
\label{sec:results}
%%%%%%%%%%%%%%%%%%%%%%%%%%%%%%%%%%%%%%%%%%%%%%%%%%%%%%%%%%%%%%%%%%%%%%%%%%

%%%%%%%%%%%%%%%%%%%%%%%%%%%%%%%%%%%%%%%%%%%%%%%%%%%%%%%%%%%%%%%%%%%%%%%%%%
\subsection{Microscopic Organization of the Low-Energy Wavefunctions}
\label{subsec:wavefunction}
%%%%%%%%%%%%%%%%%%%%%%%%%%%%%%%%%%%%%%%%%%%%%%%%%%%%%%%%%%%%%%%%%%%%%%%%%%

Before examining the covariance geometry of the low-energy manifold, it is
essential to understand the microscopic structure of the wavefunctions from
which the covariance matrices are constructed. As discussed in
Sec.~\ref{sec:model_and_method}, the basis-resolved feature vectors are
generated from the dominant computational-basis configurations of the ground
state (GS) and first excited state (FES). Consequently, the statistical
properties of the covariance matrices originate directly from the manner in
which the low-energy wavefunctions are distributed over the computational
basis.
\begin{figure*}[t]
\centering
\includegraphics[width=2.00\columnwidth]
{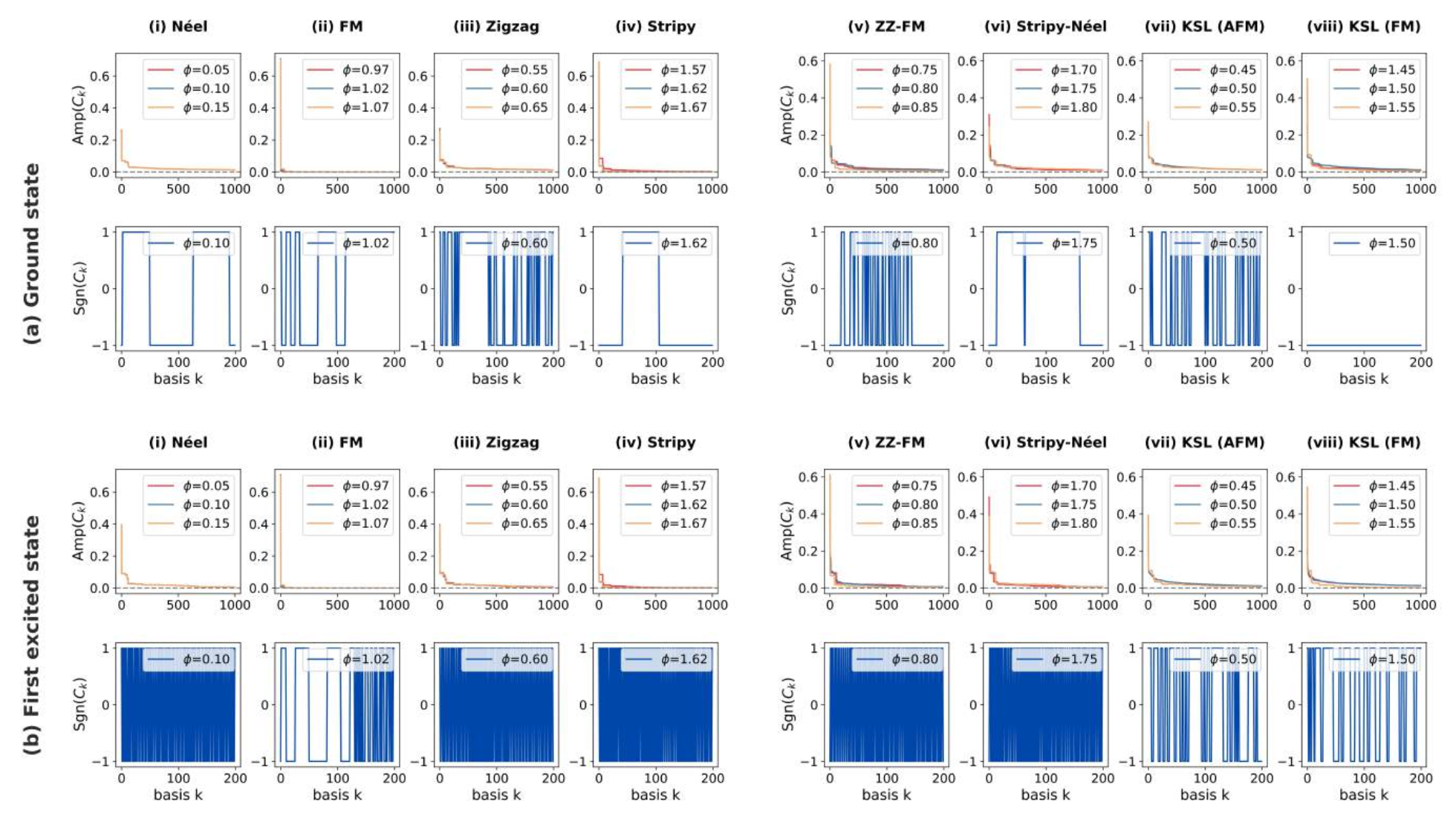}
\caption{
Expansion coefficients of the ground state (GS) and first excited state
(FES) for representative interaction parameters across the
Kitaev--Heisenberg phase diagram. For each state, the upper panel shows the
coefficient magnitudes $|C_k|$ arranged in decreasing order, while the lower
panel displays the signs of the first 200 coefficients. Only these leading signs are displayed
clearly, since the signs of all 1000 retained coefficients become visually
unresolved when plotted together. The covariance analysis nevertheless uses
all $M=1000$ retained basis configurations.
}
\label{fig:mpsc_merged}
\end{figure*}
Fig.\ref{fig:mpsc_merged} shows the expansion coefficients of the ground
and first excited states for representative interaction parameters spanning
the complete phase diagram. For each state, the coefficients are arranged in
decreasing order of their magnitude, while the relative signs of the leading
coefficients are displayed separately. Since the overall phase of an
eigenstate is arbitrary, only the relative sign structure within a given
state is physically meaningful.

A common feature of all coupling regimes is the pronounced hierarchy of the
expansion coefficients. The coefficient magnitudes decrease strongly with
their rank, indicating that the computational-basis representation of the
low-energy states is highly inhomogeneous. A limited set of configurations
carries the dominant amplitude, while a long tail of configurations
contributes with progressively smaller weight. This hierarchy provides the
microscopic basis for retaining the $M=1000$ configurations with the largest
coefficient magnitudes in the subsequent covariance construction.

Although the ranked amplitudes display a common decaying structure, their
detailed distribution changes across the phase diagram. In the magnetically
ordered N\'eel, ferromagnetic, zigzag, and stripy regimes, the leading weight
is relatively concentrated among a small number of dominant configurations.
This concentration is consistent with the presence of well-defined magnetic
patterns that organize the low-energy wavefunctions.

Near the zigzag--ferromagnetic and stripy--N\'eel phase boundaries, the
leading amplitudes are distributed more broadly among several configurations.
The wavefunction is therefore not governed by a single dominant microscopic
pattern, but reflects the competition between neighbouring magnetic orders.
The magnetic phase boundaries thus appear as regions in which the dominant
computational-basis content of the low-energy states is reorganized.

The antiferromagnetic and ferromagnetic Kitaev regimes also exhibit a broad
hierarchy of relevant basis configurations. The low-energy wavefunctions
remain compressible in the sense that the coefficients decrease with rank,
but their leading weight is distributed over a wider set of configurations
than in the conventional ordered phases. The Kitaev regimes should therefore
not be interpreted as completely delocalized states in the computational
basis; rather, they exhibit a richer participation of dominant microscopic
configurations.
\begin{figure*}[t]
\centering
\includegraphics[width=2.00\columnwidth]{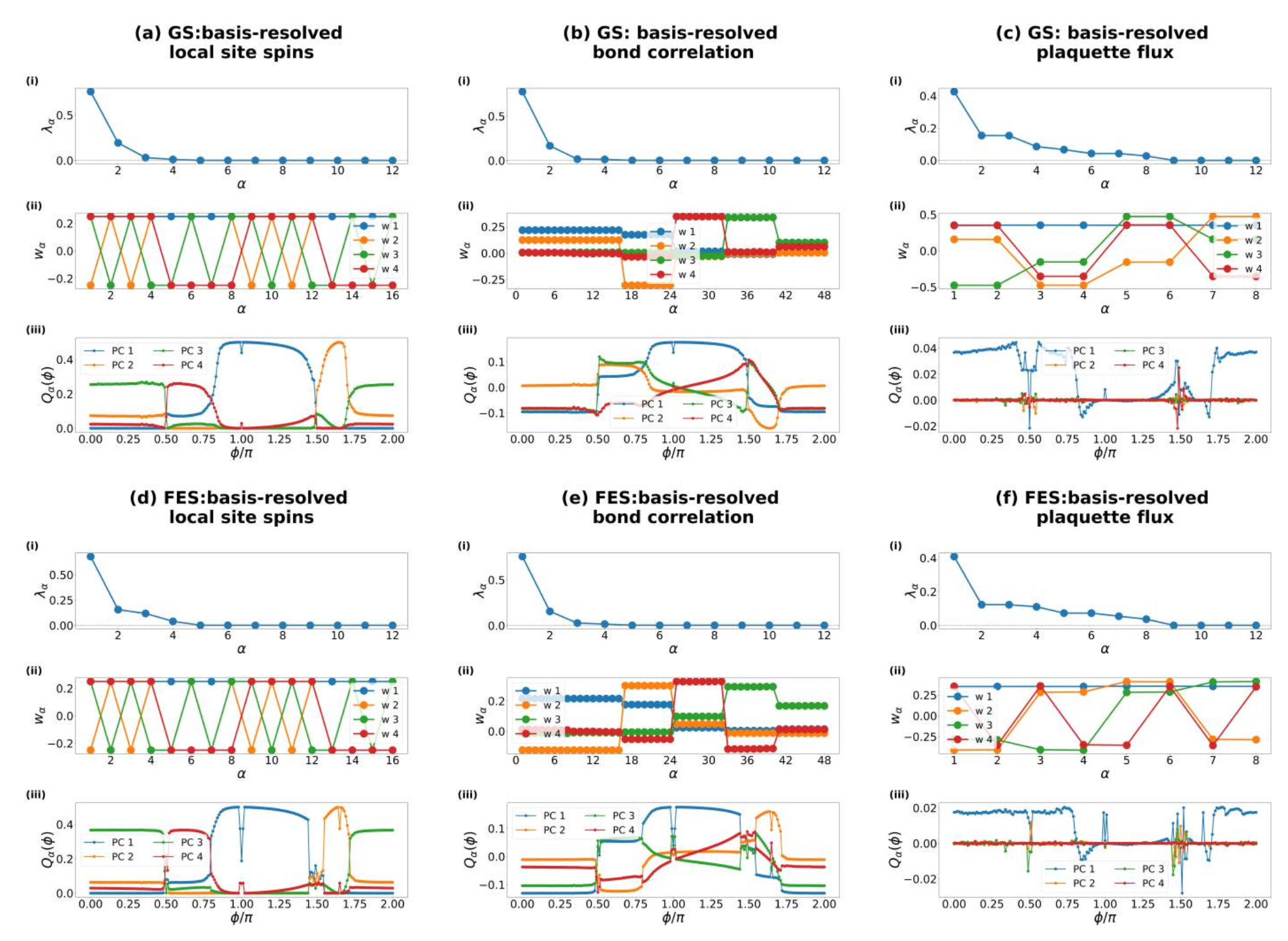}
\caption{
Principal component analysis of the basis-resolved covariance matrices for
the local-spin, spin-pair, and plaquette-flux representations. Panels (a)--(c)
correspond to the ground state, while panels (d)--(f) show the corresponding
results for the first excited state. For each observable, row (i) presents
the normalized covariance eigenvalues $\lambda_\alpha$, row (ii) the
corresponding covariance eigenvectors $w_\alpha$, and row (iii) the
quantified principal components $Q_\alpha(\phi)$ defined in
Eq.~(\ref{eqn:Qalpha}).
}
\label{fig:pca_merged}
\end{figure*}
The sign panels provide complementary information about the relative phase
organization of the important basis states. Since the eigenvectors are
obtained directly from diagonalization, their overall sign is arbitrary, and
no physical meaning is assigned to whether a given state is predominantly
positive or negative. Only the relative signs among the significant
coefficients are relevant. Moreover, sign variations associated with the
very small coefficients in the long-amplitude tail have little influence on
the wavefunction because their contributions are suppressed by their small
magnitudes.

The physically meaningful sign information is therefore carried primarily by
the leading coefficients. These signs determine how the dominant basis
configurations interfere when acted upon by off-diagonal operators such as
the bond and plaquette-flux observables. The ground and first excited states
often display similar ranked amplitude profiles while differing more visibly
in the relative-sign organization of their leading coefficients. This
indicates that low-energy excitations may involve not only a redistribution
of amplitude but also a reorganization of the interference relations among
the dominant configurations.

Fig.\ref{fig:mpsc_merged} therefore establishes two microscopic ingredients
that enter the later covariance analysis: the distribution of amplitude
among the dominant computational-basis configurations and the relative-sign
organization of those configurations. The evolution of these ingredients
across ordered phases, magnetic phase boundaries, and Kitaev regimes provides
the microscopic foundation for the changes in covariance geometry discussed
below.
%%%%%%%%%%%%%%%%%%%%%%%%%%%%%%%%%%%%%%%%%%%%%%%%%%%%%%%%%%%%%%%%%%%%%%%%%%
\subsection{Covariance Geometry of the Basis-Resolved Ensemble}
\label{subsec:covariance}
%%%%%%%%%%%%%%%%%%%%%%%%%%%%%%%%%%%%%%%%%%%%%%%%%%%%%%%%%%%%%%%%%%%%%%%%%%
Having established the microscopic organization of the low-energy
wavefunctions, we now examine how these basis-resolved configurations are
statistically organized through the covariance matrix. Fig.\ref{fig:pca_merged}
summarizes the covariance analysis for the three feature representations
introduced in Sec.~\ref{sec:model_and_method}: the local-spin, spin-pair, and
plaquette-flux features. For each representation we present the covariance
eigenvalue spectrum, the dominant covariance eigenvectors, and the quantified
principal components for both the ground state (GS) and the first excited
state (FES). Together these quantities characterize the dimensionality of the
covariance space, the microscopic structure of the dominant covariance modes,
and their evolution across the phase diagram.
The covariance spectra shown in Fig.~\ref{fig:pca_merged}(i) reveal that the
different observables possess markedly different statistical dimensionalities.
For both the GS and FES, the local-spin and spin-pair representations are
dominated by a single covariance mode, with the remaining eigenvalues
decreasing rapidly. Most of the statistical variance of these ensembles is
therefore captured by only one or two collective directions in covariance
space. In contrast, the plaquette-flux representation exhibits a broader
eigenvalue distribution, indicating that its statistical fluctuations are
distributed among several covariance modes. The covariance dimensionality is
thus observable dependent, reflecting the distinct microscopic organization
of spin, bond, and flux degrees of freedom.

The covariance eigenvectors displayed in
Fig.~\ref{fig:pca_merged}(ii) further demonstrate that the dominant covariance
modes are highly structured rather than random combinations of basis-resolved
features. The leading eigenvectors organize the retained basis configurations
into well-defined collective patterns, with characteristic sign changes and
weight redistributions occurring only at specific interaction regions. These
structured covariance modes indicate that the statistical organization of the
low-energy ensemble is governed by coherent collective fluctuations rather
than independent variations of individual basis configurations.

The quantified principal components,
Fig.~\ref{fig:pca_merged}(iii), reveal how the importance of these covariance
modes evolves throughout the phase diagram. Since
$Q_\alpha(\phi)$ measures the average magnitude of the projection of the
basis-resolved samples onto the $\alpha$th covariance mode, it directly
quantifies the statistical activity of that mode at a given interaction
parameter. The dominant covariance mode changes systematically as $\phi$ is
varied, with different principal components becoming active in different
regions of the phase diagram. Rather than being uniformly important, each
covariance mode characterizes a specific statistical organization of the
basis-resolved ensemble, and the evolution of $Q_\alpha(\phi)$ tracks the
continuous reorganization of these collective statistical patterns across the
competing magnetic phases and the Kitaev regimes.

A comparison between the GS and FES demonstrates that the overall covariance
geometry is remarkably robust. The eigenvalue spectra, covariance
eigenvectors, and quantified principal components exhibit closely related
structures for both low-energy states, indicating that the dominant
statistical organization is inherited by the low-energy manifold rather than
being a property of an individual eigenstate. Fig.~\ref{fig:pca_merged}.(a).(iii) \& Fig.~\ref{fig:pca_merged}.(d).(iii) and Fig.~\ref{fig:pca_merged}.(b).(iii) \& Fig.~\ref{fig:pca_merged}.(e).(iii) are similar to the results shown in Fig.4.(b) and Fig.5.(a) respectively in the presented work by Gotfryd \cite{gotfryd2017phase}. This robustness suggests that
the covariance modes capture intrinsic collective characteristics of the
underlying quantum phases.

Fig.\ref{fig:pca_merged} therefore establishes that the basis-resolved
ensemble is organized by a small number of collective covariance modes whose
number, structure, and activity depend strongly on the chosen observable.
The local-spin and spin-pair representations are governed by a few dominant
collective modes, whereas the plaquette-flux representation exhibits a richer
covariance structure. These covariance modes provide the statistical
foundation for identifying the distinct quantum phases and their boundaries
without introducing explicit order parameters.

%%%%%%%%%%%%%%%%%%%%%%%%%%%%%%%%%%%%%%%%%%%%%%%%%%%%%%%%%%%%%%%%%%%%%%%%%%
\subsection{Covariance Geometry within the Ordered Phases}
\label{subsec:ordered_phases}
%%%%%%%%%%%%%%%%%%%%%%%%%%%%%%%%%%%%%%%%%%%%%%%%%%%%%%%%%%%%%%%%%%%%%%%%%%
\begin{figure}[t]
\centering
\includegraphics[width=\columnwidth]{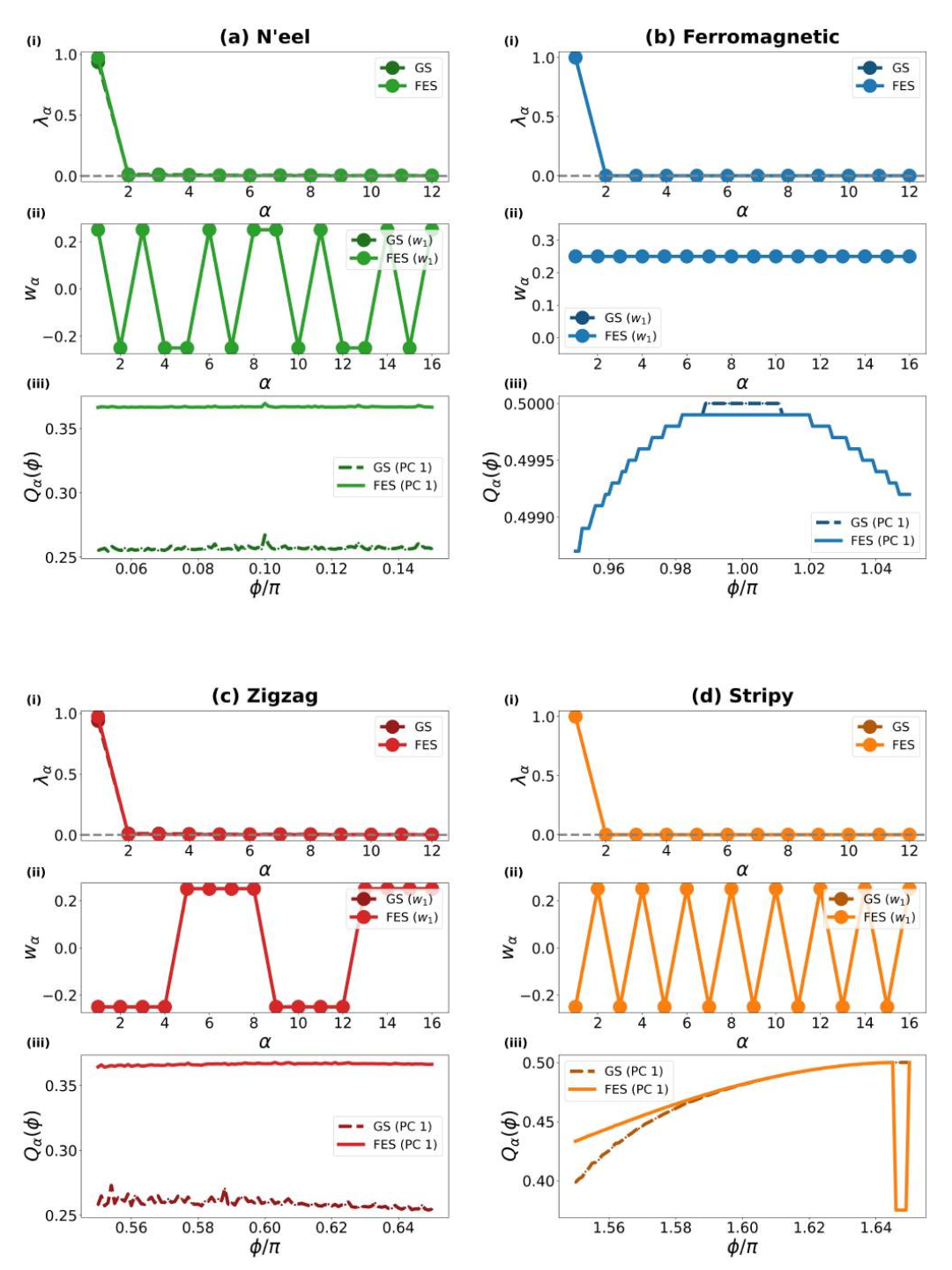}
\caption{
Principal component analysis of the basis-resolved local-spin covariance
matrix within the four ordered phases. Panels (a)--(d) correspond to the
N\'eel, ferromagnetic, zigzag, and stripy phases, respectively. For each
phase, row (i) shows the normalized covariance eigenvalue spectrum, row (ii)
the dominant covariance eigenvector, and row (iii) the quantified principal
component $Q_1(\phi)$ for both the ground state (GS) and the first excited
state (FES).
}
\label{fig:ordered_spin}
\end{figure}
Fig.\ref{fig:ordered_spin} investigates the covariance geometry deep
inside the four magnetically ordered phases using the basis-resolved
local-spin representation. While Fig.~\ref{fig:pca_merged} established the
global evolution of the covariance modes across the entire phase diagram, the
present analysis focuses on the internal statistical structure of each
ordered phase. Representative coupling regions are selected sufficiently away
from the phase boundaries so that the covariance properties reflect the
intrinsic characteristics of the corresponding magnetic order.

The covariance spectra shown in Fig.~\ref{fig:ordered_spin}(i) exhibit a
remarkably simple structure. In all four ordered phases, nearly the entire
covariance variance is captured by the leading eigenvalue, while the
remaining eigenvalues are negligibly small. The covariance matrix therefore
becomes effectively rank one throughout the ordered regions, demonstrating
that the statistical fluctuations of the basis-resolved ensemble are governed
by a single collective covariance mode. This observation strengthens the
global picture presented in Fig.~\ref{fig:pca_merged}, where the ordered
phases were already found to be dominated by the leading principal component.
Deep inside each ordered phase, this dominance becomes essentially complete.

Although all ordered phases possess an effectively one-dimensional covariance
geometry, the corresponding dominant covariance eigenvectors exhibit distinct
structures, as shown in Fig.~\ref{fig:ordered_spin}(ii). The ferromagnetic
phase is characterized by an almost uniform eigenvector, reflecting the
homogeneous nature of the underlying spin alignment. In contrast, the
N\'eel, zigzag, and stripy phases display characteristic alternating or
block-like sign patterns that encode the symmetry of their respective
magnetic orders. Thus, while the covariance dimensionality is universal among
the ordered phases, the dominant covariance mode itself provides a unique
statistical fingerprint of each magnetic state.

Further insight is obtained from the quantified principal component
$Q_1(\phi)$ shown in Fig.~\ref{fig:ordered_spin}(iii). Throughout each
ordered phase, $Q_1(\phi)$ remains nearly constant, indicating that the same
covariance mode governs the statistical organization of the retained
basis-resolved ensemble over the entire phase. Only weak variations are
observed, primarily near the edges of the selected coupling intervals where
the influence of neighbouring phases begins to emerge. These results indicate
that the internal covariance geometry of an ordered phase is remarkably
stable and does not undergo significant reorganization away from the phase
boundaries.

A comparison between the GS and FES further demonstrates the robustness of
this covariance structure. The covariance spectra, dominant eigenvectors, and
quantified principal components nearly overlap for the two low-energy states
in all four ordered phases. The dominant covariance mode is therefore not a
special property of the ground state alone but is inherited by the
low-energy manifold. This robustness is consistent with the wavefunction
analysis presented in Fig.~\ref{fig:mpsc_merged}, where both the GS and FES
were found to possess similar hierarchical basis representations, differing
only through relatively minor redistributions of amplitude and sign among the
leading basis configurations.

Taken together, Figs.~\ref{fig:mpsc_merged},
\ref{fig:pca_merged}, and \ref{fig:ordered_spin} reveal a consistent
hierarchy of statistical organization. Fig.\ref{fig:mpsc_merged}
established that the ordered phases are represented by highly concentrated
wavefunctions in the computational basis. Fig.\ref{fig:pca_merged}
demonstrated that these basis-resolved ensembles are globally described by a
small number of covariance modes. The present figure shows that, once the
system is deep inside an ordered phase, this statistical organization
collapses to an essentially one-dimensional covariance geometry whose unique
dominant covariance mode serves as a compact statistical fingerprint of the
underlying magnetic order.

%%%%%%%%%%%%%%%%%%%%%%%%%%%%%%%%%%%%%%%%%%%%%%%%%%%%%%%%%%%%%%%%%%%%%%%%%%
\subsection{Covariance Reorganization Near Magnetic Phase Boundaries}
\label{subsec:phase_boundary}
%%%%%%%%%%%%%%%%%%%%%%%%%%%%%%%%%%%%%%%%%%%%%%%%%%%%%%%%%%%%%%%%%%%%%%%%%%

\begin{figure}[t]
\centering
\includegraphics[width=\columnwidth]{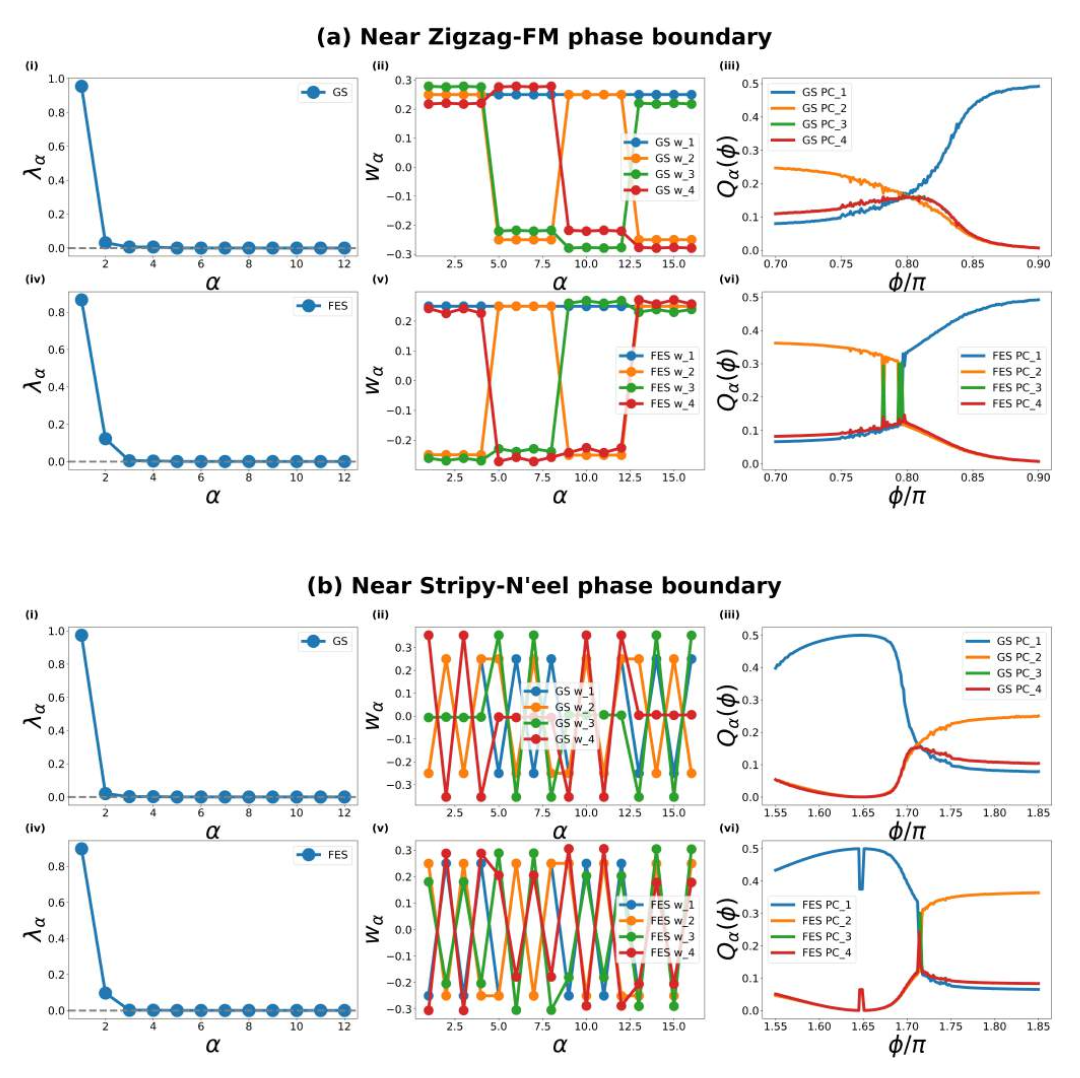}
\caption{
Covariance analysis of the basis-resolved local-spin representation near the
(a) zigzag--FM and (b) stripy--N\'eel phase boundaries. For both the ground
state (GS) and first excited state (FES), rows (i) and (iv) show the
normalized covariance eigenvalues, rows (ii) and (v) the dominant covariance
eigenvectors, and rows (iii) and (vi) the quantified principal components
$Q_\alpha(\phi)$.
}
\label{fig:phase_boundary_spin}
\end{figure}

The ordered phases discussed above are characterized by a single dominant
covariance mode with a nearly constant statistical activity throughout the
phase. An important question is therefore how this simple covariance
structure evolves as the system approaches a magnetic phase transition.
Fig.\ref{fig:phase_boundary_spin} addresses this question by examining the
basis-resolved local-spin covariance in the vicinity of the zigzag--FM and
stripy--N\'eel phase boundaries.

The covariance spectra shown in
Fig.~\ref{fig:phase_boundary_spin}(i) and (iv) remain strongly dominated by
the leading eigenvalue throughout both phase-boundary regions. In contrast to
what might be expected for competing magnetic states, no substantial increase
in the number of important covariance modes is observed. The covariance
matrix therefore remains effectively low dimensional even in the vicinity of
the phase transitions. The statistical complexity of the basis-resolved
ensemble is thus not associated with an increase in covariance dimensionality.

The covariance eigenvectors, shown in
Fig.~\ref{fig:phase_boundary_spin}(ii) and (v), exhibit a considerably richer
structure than that found deep inside the ordered phases. Whereas the ordered
phases possess simple phase-specific covariance patterns
(Fig.~\ref{fig:ordered_spin}), the phase-boundary regions display pronounced
changes in the distribution of the covariance weights among the retained
basis configurations. These reorganizations reflect the competition between
the neighbouring magnetic orders and indicate that the dominant statistical
patterns of the basis-resolved ensemble become increasingly mixed as the
phase boundary is approached.

The most direct signature of this competition is provided by the quantified
principal components shown in
Fig.~\ref{fig:phase_boundary_spin}(iii) and (vi). Away from the transition,
one principal component clearly dominates, consistent with the one-dimensional
covariance geometry established in Fig.~\ref{fig:ordered_spin}. As the phase
boundary is approached, however, the dominant covariance mode gradually loses
statistical weight while a second principal component simultaneously becomes
more important. The transition is therefore characterized by a continuous
transfer of statistical importance between competing covariance modes rather
than by the emergence of additional statistically significant directions.
Higher-order principal components become appreciable only within a narrow
region around the transition, reflecting enhanced competition between the
underlying microscopic configurations.

The behaviour differs quantitatively for the two phase boundaries. The
zigzag--FM transition exhibits a comparatively broad redistribution of the
principal-component activity, whereas the stripy--N\'eel boundary displays a
much sharper exchange between the leading covariance modes. The covariance
analysis therefore distinguishes not only the neighbouring magnetic phases
but also the different statistical character of their phase boundaries.

A comparison between the GS and FES again reveals a remarkable robustness of
the covariance geometry. Although minor quantitative differences are visible,
both low-energy states exhibit the same dominant covariance spectra, similar
covariance eigenvectors, and nearly identical evolution of the quantified
principal components. The statistical reorganization associated with the
phase transition is therefore a property of the low-energy manifold rather
than of an individual eigenstate.

Taken together with Figs.~\ref{fig:mpsc_merged},
\ref{fig:pca_merged}, and \ref{fig:ordered_spin}, the present results
complete a coherent picture of the covariance organization across the phase
diagram. Deep inside an ordered phase, the basis-resolved ensemble is
governed by a single dominant covariance mode. Near a phase boundary, the
system does not generate a larger number of statistically important
directions; instead, the statistical dominance is transferred from one
collective covariance mode to another. The magnetic phase transition is thus
manifested as a reorganization of the dominant covariance patterns of the
basis-resolved ensemble while preserving its intrinsically low-dimensional
covariance structure.

%%%%%%%%%%%%%%%%%%%%%%%%%%%%%%%%%%%%%%%%%%%%%%%%%%%%%%%%%%%%%%%%%%%%%%%%%%
\subsection{Covariance Geometry Near the Kitaev Points}
\label{subsec:kitaev_points}
%%%%%%%%%%%%%%%%%%%%%%%%%%%%%%%%%%%%%%%%%%%%%%%%%%%%%%%%%%%%%%%%%%%%%%%%%%

\begin{figure}[t]
\centering
\includegraphics[width=\columnwidth]{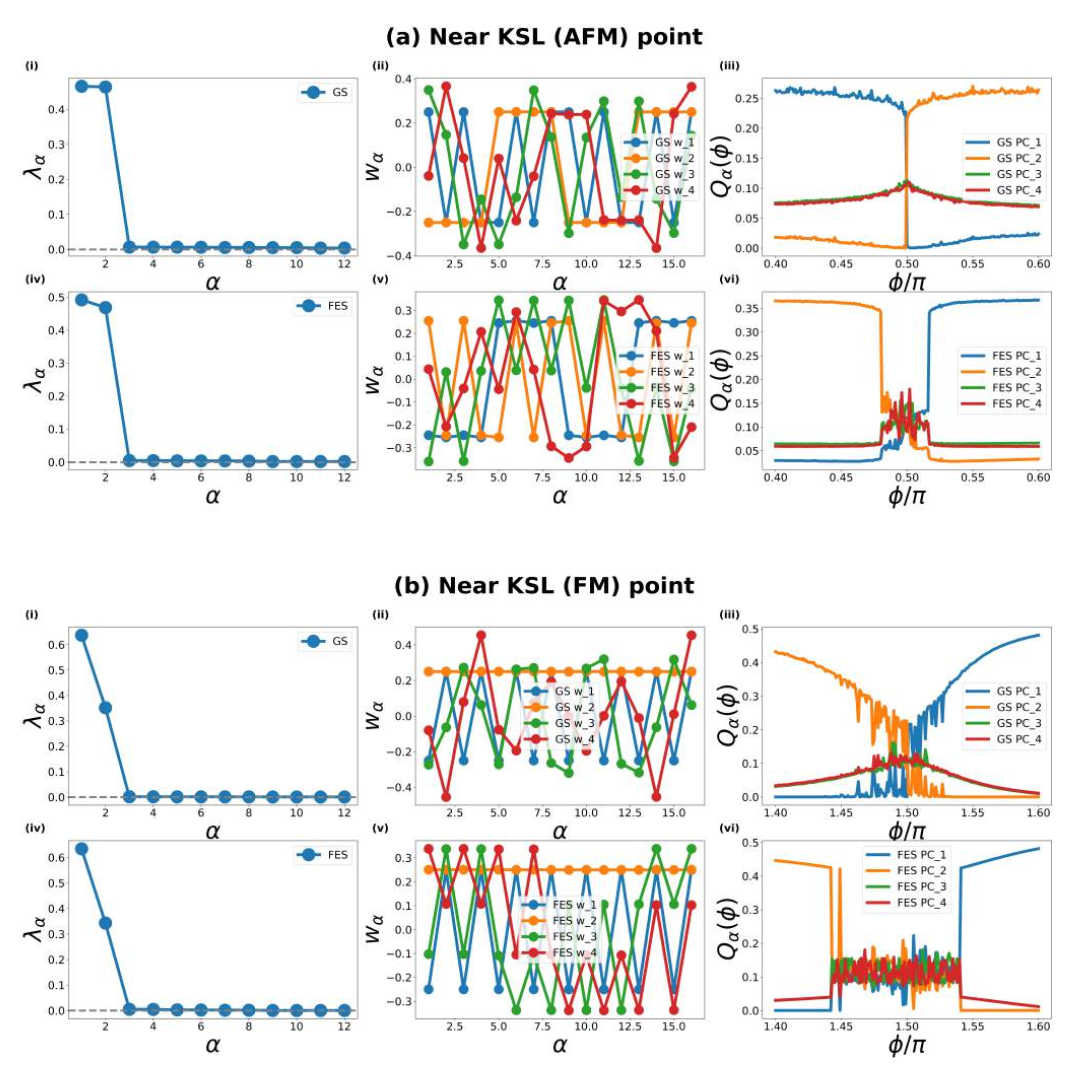}
\caption{
Covariance analysis of the basis-resolved local-spin representation near the
(a) AFM Kitaev point and (b) FM Kitaev point. For both the ground state (GS)
and first excited state (FES), rows (i) and (iv) show the normalized
covariance eigenvalues, rows (ii) and (v) the dominant covariance
eigenvectors, and rows (iii) and (vi) the quantified principal components
$Q_\alpha(\phi)$.
}
\label{fig:kitaev_spin}
\end{figure}

The previous section demonstrated that conventional magnetic phase
boundaries are characterized by a transfer of statistical dominance between
competing covariance modes while the covariance matrix itself remains
essentially one dimensional. We now examine whether the same statistical
picture remains valid in the vicinity of the antiferromagnetic (AFM) and
ferromagnetic (FM) Kitaev points. The corresponding covariance analysis of
the basis-resolved local-spin representation is shown in
Fig.~\ref{fig:kitaev_spin}.

A qualitative change in the covariance geometry is immediately evident from
the covariance spectra shown in
Fig.~\ref{fig:kitaev_spin}(i) and (iv). Unlike the ordered phases
(Fig.~\ref{fig:ordered_spin}) and the conventional magnetic phase boundaries
(Fig.~\ref{fig:phase_boundary_spin}), where the leading covariance mode
accounts for nearly the entire statistical variance, the Kitaev regions
display a substantially broader covariance modes. The leading
eigenvalue no longer dominates the covariance matrix, while the second
covariance mode carries a comparable fraction of the statistical variance.
The covariance geometry therefore becomes intrinsically multidimensional in
the vicinity of both the AFM and FM Kitaev points.

The covariance eigenvectors shown in
Fig.~\ref{fig:kitaev_spin}(ii) and (v) likewise differ markedly from those
observed in the ordered phases. Instead of exhibiting the simple alternating
or block-like structures associated with conventional magnetic order, the
covariance weights are distributed among several competing collective
patterns. No single covariance eigenvector uniquely characterizes the
statistical organization of the retained basis configurations. This behavior
is consistent with the wavefunction analysis of
Fig.~\ref{fig:mpsc_merged}, where the Kitaev states were found to distribute
their probability over a substantially larger number of computational-basis
configurations than the ordered phases.

The quantified principal components,
Fig.~\ref{fig:kitaev_spin}(iii) and (vi), provide the clearest evidence for
this multidimensional covariance structure. As the interaction parameter
approaches the Kitaev point, the dominant principal component rapidly loses
its statistical weight while a second principal component becomes equally
important. Simultaneously, higher-order principal components also acquire
appreciable magnitude within a narrow interval surrounding the Kitaev point.
Unlike the conventional magnetic boundaries discussed in
Fig.~\ref{fig:phase_boundary_spin}, where the transition is primarily
described by a transfer of statistical dominance between two competing
covariance modes, the Kitaev regions exhibit the simultaneous participation
of several statistically significant covariance modes.

The covariance organization is again remarkably similar for the GS and FES.
Both low-energy states exhibit comparable covariance spectra, closely related
covariance eigenvectors, and nearly identical evolution of the quantified
principal components. The multidimensional covariance geometry is therefore
not a property of an individual eigenstate but a robust characteristic of
the low-energy manifold in the vicinity of the Kitaev points.

Taken together, Figs.~\ref{fig:ordered_spin},
\ref{fig:phase_boundary_spin}, and \ref{fig:kitaev_spin} establish three
distinct statistical regimes of the basis-resolved covariance geometry.
Deep inside the ordered phases, the covariance matrix is effectively
one-dimensional and is governed by a unique dominant covariance mode.
Conventional magnetic phase boundaries preserve this low-dimensional
structure while transferring statistical dominance between competing
principal components. In contrast, the Kitaev regions display an intrinsic
coexistence of multiple statistically significant covariance modes,
reflecting a fundamentally richer covariance geometry. Combined with the
broader computational-basis participation observed in
Fig.~\ref{fig:mpsc_merged}, these results demonstrate that the Kitaev states
are distinguished not merely by the redistribution of wavefunction
amplitudes but also by the emergence of genuinely multidimensional
statistical correlations in the basis-resolved ensemble.

%%%%%%%%%%%%%%%%%%%%%%%%%%%%%%%%%%%%%%%%%%%%%%%%%%%%%%%%%%%%%%%%%%%%%%%%%%
\subsection{Flux Covariance Geometry Near the Kitaev Points}
\label{subsec:flux_kitaev}
%%%%%%%%%%%%%%%%%%%%%%%%%%%%%%%%%%%%%%%%%%%%%%%%%%%%%%%%%%%%%%%%%%%%%%%%%%

\begin{figure}[t]
\centering
\includegraphics[width=\columnwidth]{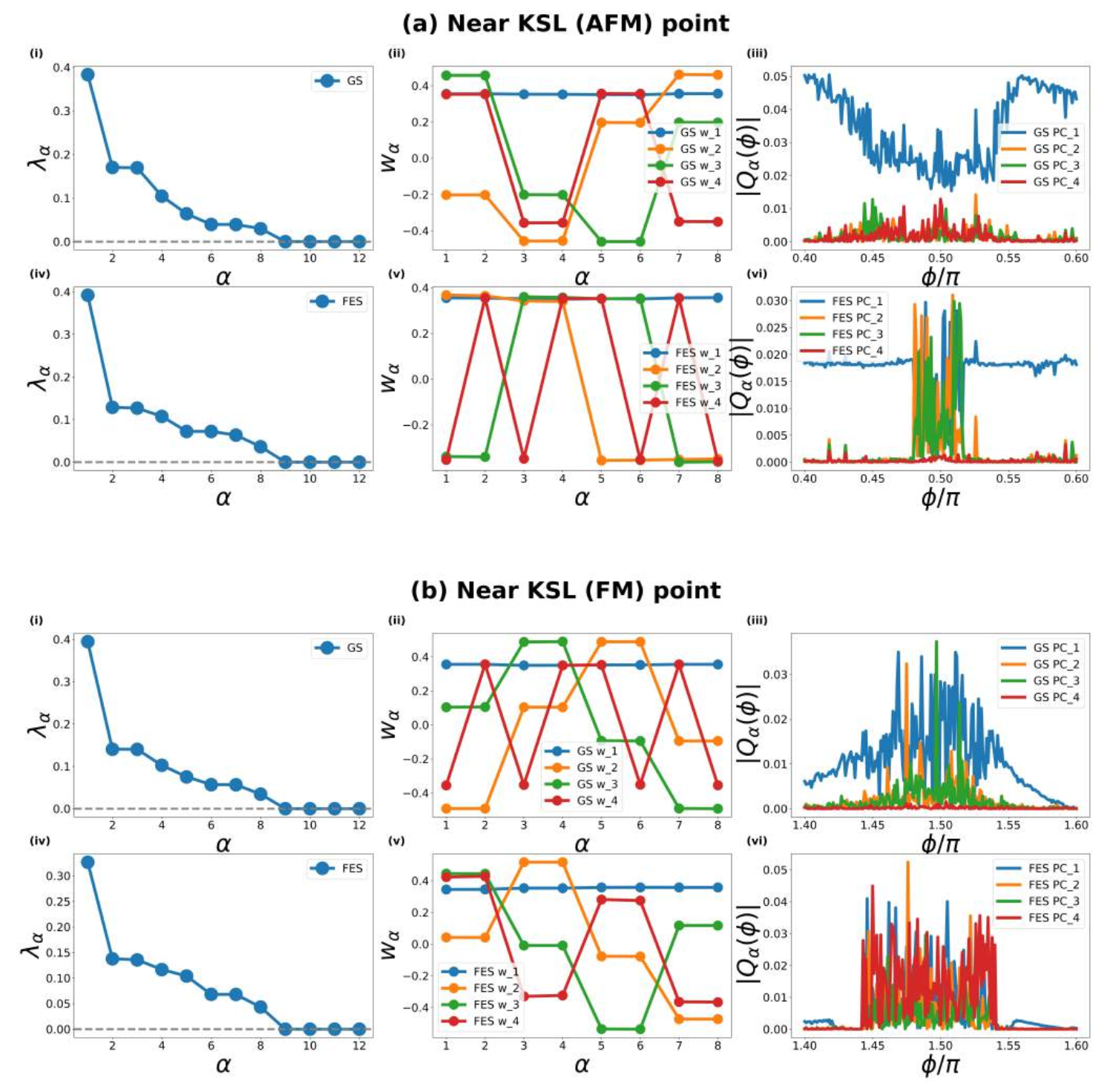}
\caption{
Principal component analysis of the basis-resolved plaquette-flux covariance
matrix near the AFM and FM Kitaev points. Panels (a) and (b) correspond to
the two Kitaev regions. For both the ground state (GS) and first excited
state (FES), rows (i) and (iv) show the normalized covariance eigenvalue
spectra, rows (ii) and (v) the dominant covariance eigenvectors, and rows
(iii) and (vi) the quantified principal components $Q_\alpha(\phi)$.
}
\label{fig:kitaev_flux}
\end{figure}

The local-spin representation discussed in the previous section revealed that
the covariance geometry near the Kitaev points becomes intrinsically
multidimensional through the simultaneous participation of several principal
components. An important question is whether this statistical organization is
a universal property of the wavefunction or depends on the physical
observable used to construct the covariance matrix. To address this issue,
Fig.~\ref{fig:kitaev_flux} presents the corresponding analysis for the
basis-resolved plaquette-flux representation.

A striking difference from the local-spin representation is immediately
apparent in the covariance spectra shown in
Fig.~\ref{fig:kitaev_flux}(i) and (iv). Instead of being dominated by one or
two leading covariance modes, the plaquette-flux covariance distributes its
statistical variance over a much larger number of principal components. The
eigenvalue spectrum therefore decays much more gradually, indicating that the
flux representation possesses a substantially higher intrinsic covariance
dimensionality than the corresponding local-spin representation.

The covariance eigenvectors shown in
Fig.~\ref{fig:kitaev_flux}(ii) and (v) also exhibit a qualitatively different
organization. Unlike the phase-specific patterns observed for the ordered
phases or the relatively simple reorganizations found in the local-spin
representation near the Kitaev points, the flux covariance weights are
distributed among several competing collective patterns. No individual
covariance mode dominates the statistical organization of the retained basis
ensemble, reflecting the intrinsically collective nature of the plaquette
degrees of freedom.

This richer covariance structure is further demonstrated by the quantified
principal components shown in
Fig.~\ref{fig:kitaev_flux}(iii) and (vi). In contrast to the local-spin
representation, where one or two principal components remain dominant even in
the vicinity of the Kitaev points, the plaquette-flux representation exhibits
the simultaneous activity of several covariance modes over the entire Kitaev
region. The statistical information is therefore fragmented among many
collective directions rather than being concentrated in a few dominant
principal components. The absolute magnitude of $Q_\alpha(\phi)$ is much
smaller than for the local-spin representation because of the different
normalization and physical nature of the plaquette-flux operators; the
relevant information is contained in the relative distribution of the
principal-component activity rather than in its absolute scale.

The covariance organization remains remarkably similar for the GS and FES.
Although small quantitative differences are visible, both low-energy states
display nearly identical covariance spectra and comparable distributions of
principal-component activity. The higher-dimensional covariance geometry is
therefore a robust property of the low-energy manifold rather than a feature
of a particular eigenstate.

The comparison between Figs.~\ref{fig:kitaev_spin} and
\ref{fig:kitaev_flux} demonstrates that the covariance geometry depends
strongly on the physical observable used to represent the low-energy
wavefunctions. The local-spin representation retains a partially
low-dimensional statistical organization even near the Kitaev points,
whereas the plaquette-flux representation distributes the covariance variance
over many collective modes. The emergent covariance geometry is therefore not
a universal property of the wavefunction alone but also reflects the
underlying physical correlations encoded by the chosen observable.

%%%%%%%%%%%%%%%%%%%%%%%%%%%%%%%%%%%%%%%%%%%%%%%%%%%%%%%%%%%%%%%%%%%%%%%%%%
\subsection{Quantifying the Covariance Geometry}
\label{subsec:covariance_quantification}
%%%%%%%%%%%%%%%%%%%%%%%%%%%%%%%%%%%%%%%%%%%%%%%%%%%%%%%%%%%%%%%%%%%%%%%%%%
\begin{table}[t]
\centering
\caption{Shannon's entropy analysis for GS and FES.}
\label{tab:shannon_entropy}
\begin{tabular}{l ccc ccc}
\toprule
\multicolumn{1}{c}{\textbf{$\mathcal{S}$}} &
\multicolumn{3}{c}{\textbf{GS}} & \multicolumn{3}{c}{\textbf{FES}} \\
\hline
\multicolumn{1}{c}{$\phi$ (\textbf{Regime})} & Site & Bond & Plaq & Site & Bond & Plaq \\
\hline
0.05--0.15 & 0.2932 & 0.5933 & 1.6736 & 0.1372 & 0.3999 & 1.7636 \\
0.55--0.65 & 0.2831 & 0.5908 & 1.6917 & 0.1372 & 0.4074 & 1.7671 \\
0.95--1.05 & 0.0000 & 0.0045 & 1.7016 & 0.0000 & 0.0043 & 1.7011 \\
1.55--1.65 & 0.0071 & 0.1136 & 1.7526 & 0.0097 & 0.1165 & 1.7916 \\
\hline
0.70-0.90 & 0.2245 & 0.2951 & 1.4828 & 0.4367 & 0.5241 & 1.7204 \\
1.55-1.85 & 0.1367 & 0.2685 & 1.6073 & 0.3544 & 0.5046 & 1.7519 \\
\hline
0.40--0.60 & 0.8984 & 1.0777 & 1.7412 & 0.8379 & 0.9479 & 1.8082 \\
1.40-1.60 & 0.7008 & 0.8032 & 1.7867 & 0.7626 & 0.8007 & 1.8969 \\
\hline
\end{tabular}
\end{table}
The covariance analyses presented in
Figs.~\ref{fig:pca_merged}--\ref{fig:kitaev_flux}
establish the statistical organization of the basis-resolved ensemble
through the covariance spectra and the quantified principal components.
To quantify these observations, we evaluate two complementary measures
derived from the covariance eigenvalue distribution: Shannon entropy
($\mathcal{S}$) and the participation ratio (PR). These quantities
measure, respectively, the statistical dispersion of the covariance
variance and the effective number of statistically significant covariance
modes.

For a normalized covariance spectrum

\begin{equation}
p_\alpha=\frac{\lambda_\alpha^2}
{\sum_\beta\lambda_\beta^2},
\end{equation}

the Shannon entropy is defined as

\begin{equation}
\mathcal S
=
-\sum_\alpha
p_\alpha\ln p_\alpha,
\end{equation}

while the participation ratio is

\begin{equation}
\mathrm{PR}
=
\frac{\left(\sum_\alpha p_\alpha\right)^2}
{\sum_\alpha p_\alpha^2}.
\end{equation}

The Shannon entropy measures how uniformly the covariance variance is
distributed among the principal components, whereas the participation
ratio estimates the effective number of statistically active covariance
modes. Small values of $\mathcal S$ and
$\mathrm{PR}\approx1$ indicate an essentially one-dimensional covariance
geometry, while larger values imply progressively richer and more
multidimensional covariance structures.

Table~\ref{tab:shannon_entropy} summarizes the Shannon entropy for the
local-spin, bond-correlation, and plaquette-flux representations in the
ordered phases, magnetic phase boundaries, and Kitaev regions for both
the ground state (GS) and first excited state (FES).

%%%%%%%%%%%%%%%%%%%%%%%%%%%%%%%%%%%%%%%%%%%%%%%%%%%%%%%%%%%%%%%%%%%%%%
%% TABLE II
%%%%%%%%%%%%%%%%%%%%%%%%%%%%%%%%%%%%%%%%%%%%%%%%%%%%%%%%%%%%%%%%%%%%%%

%\begin{table*}[t]
%\centering
%\caption{Shannon entropy ($\mathcal S$) computed from the covariance
%eigenvalue spectrum for the local-spin (Site), bond-correlation (Bond),
%and plaquette-flux (Plaq) representations.}
%\label{tab:shannon_entropy}

% <<<<<<<<<<<<<<<<<<<
% Paste your complete Table II here exactly as you already prepared it.
% <<<<<<<<<<<<<<<<<<<
%\end{table*}

The numerical values quantitatively confirm the covariance geometries
identified in the preceding figures. Deep inside the ordered phases,
the local-spin and bond representations exhibit extremely small entropy,
consistent with the rank-one covariance structure observed in
Fig.~\ref{fig:ordered_spin}. The entropy increases near the magnetic
phase boundaries, reflecting the redistribution of covariance variance
between competing principal components, and reaches its largest values
near the Kitaev regions where several covariance modes become
simultaneously active. In contrast, the plaquette-flux representation
maintains comparatively large entropy throughout the phase diagram,
indicating that its covariance variance is intrinsically distributed
among many collective modes.

To complement this information, the effective number of statistically
important covariance modes is quantified through the participation
ratio and summarized in Table.\ref{tab:participation_ratio}.

\begin{table}[t]
\centering
\caption{Participation Ratio for GS and FES.}
\label{tab:participation_ratio}
\begin{tabular}{l ccc ccc}
\toprule
\multicolumn{1}{c}{\textbf{PR}} &
\multicolumn{3}{c}{\textbf{GS}} & \multicolumn{3}{c}{\textbf{FES}} \\
\hline
\multicolumn{1}{c}{$\phi$ (\textbf{Regime})} & Site & Bond & Plaq & Site & Bond & Plaq \\
\hline
0.05--0.15 & 1.1080 & 1.2831 & 3.9850 & 1.0414 & 1.1607 & 4.3379 \\
0.55--0.65 & 1.1029 & 1.2813 & 4.0594 & 1.0414 & 1.1634 & 4.3653 \\
0.95--1.05 & 1.0000 & 1.0048 & 3.8160 & 1.0000 & 1.008 & 3.8124 \\
1.55--1.65 & 1.0014 & 1.0349 & 4.2736 & 1.0020 & 1.0354 & 4.5855 \\
\hline
0.70-0.90 & 1.0948 & 1.1253 & 3.1823 & 1.3015 & 1.3412 & 4.0026 \\
1.55-1.85 & 1.0525 & 1.1121 & 3.6207 & 1.2247 & 1.3136 & 4.2485 \\
\hline
0.40--0.60 & 2.1621 & 2.2335 & 4.4724 & 2.1029 & 2.0645 & 4.6731 \\
1.40-1.60 & 1.8728 & 1.8481 & 4.5750 & 1.9089 & 1.8427 & 5.5659 \\
\hline
\end{tabular}
\end{table}
The participation ratio provides a geometric interpretation of the
covariance spectrum. For the local-spin and bond representations,
$\mathrm{PR}\approx1$ throughout the ordered phases confirms that a
single covariance mode captures nearly all statistical fluctuations.
Near the conventional magnetic phase boundaries, the participation ratio
increases only moderately, demonstrating that the covariance geometry
remains effectively low dimensional despite the transfer of statistical
dominance between competing principal components
(Fig.~\ref{fig:phase_boundary_spin}). In contrast, the participation
ratio increases substantially near the Kitaev regions, quantitatively
establishing the emergence of multidimensional covariance geometry
observed in Fig.~\ref{fig:kitaev_spin}.

The plaquette-flux representation exhibits a qualitatively different
behavior. Its participation ratio remains consistently larger than that
of the local-spin and bond representations over the entire phase
diagram, confirming that the flux covariance is intrinsically
high-dimensional. Together, Tables~\ref{tab:shannon_entropy} and
\ref{tab:participation_ratio} provide quantitative validation of the
entire covariance analysis developed in this work, confirming the
systematic evolution from one-dimensional covariance geometry in the
ordered phases, through covariance-mode competition at conventional
magnetic boundaries, to genuinely multidimensional covariance
organization in the vicinity of the Kitaev points.

%%%%%%%%%%%%%%%%%%%%%%%%%%%%%%%%%%%%%%%%%%%%%%%%%%%%%%%%%%%%%%%%%%%%%%%%%%
\section{Conclusion and Outlook}
%%%%%%%%%%%%%%%%%%%%%%%%%%%%%%%%%%%%%%%%%%%%%%%%%%%%%%%%%%%%%%%%%%%%%%%%%%

In this work, we have developed a basis-resolved covariance framework for
investigating the microscopic organization of low-energy many-body
wavefunctions in the spin-$1/2$ Kitaev--Heisenberg model. Unlike conventional
PCA approaches that analyze classical snapshots, spin configurations, or
individual expectation values, the present method performs the statistical
analysis directly on basis-resolved many-body wavefunctions represented in
different physical operator spaces. Local-spin, bond-correlation, and
plaquette-flux operators are used to construct covariance matrices, whose
dominant collective fluctuation channels are identified through Principal
Component Analysis (PCA). This provides a compact statistical description of
the geometry of the low-energy manifold without relying on explicit order
parameters.

Our analysis reveals a clear hierarchy in the covariance geometry across the
phase diagram. The conventional N\'eel, ferromagnetic, zigzag, and stripy
phases are characterized by an essentially one-dimensional covariance
manifold, where nearly the entire covariance variance is captured by a single
principal mode. At the conventional magnetic phase boundaries, the covariance
geometry remains low dimensional, but the statistical weight is redistributed
between competing covariance modes, reflecting the continuous reorganization
of the underlying wavefunctions. In contrast, the Kitaev regimes exhibit
intrinsically multidimensional covariance geometry, where several principal
components contribute simultaneously to the statistical organization of the
low-energy states. Shannon entropy and the participation ratio provide
quantitative confirmation of this systematic evolution from ordered phases to
phase boundaries and finally to the Kitaev regions.

A central outcome of this work is that the covariance geometry is not an
intrinsic property of the quantum wavefunction alone, but depends
fundamentally on the physical observables used to represent it. While the
local-spin and bond representations retain relatively low-dimensional
covariance structures except near the Kitaev regimes, the plaquette-flux
representation exhibits a substantially richer covariance organization,
remaining intrinsically multidimensional throughout the phase diagram. The
same many-body wavefunction therefore gives rise to distinct covariance
geometries in different operator spaces, demonstrating that complementary
physical observables reveal different aspects of the underlying quantum
correlations.

More broadly, the present work establishes basis-resolved covariance geometry
as a complementary framework for characterizing many-body quantum states. By
focusing on the statistical organization of low-energy wavefunctions rather
than conventional symmetry-breaking order parameters, the method naturally
extends to systems where the appropriate order parameter is unknown or absent.
Consequently, the framework is directly applicable to a broad class of
strongly correlated quantum systems, including frustrated quantum magnets,
interacting fermionic models, tensor-network wavefunctions, quantum Monte
Carlo generated states, and variational quantum algorithms.

We anticipate that covariance geometry can provide a unified statistical
language for exploring emergent quantum phases and collective organization in
many-body systems. Beyond the Kitaev--Heisenberg model, the present approach
opens new possibilities for studying quantum states through the geometry of
their microscopic wavefunction fluctuations, thereby providing a bridge
between basis-level quantum information and emergent collective behavior in
strongly correlated matter.

We hope that basis-resolved covariance geometry will provide a general statistical framework for studying many-body wavefunctions in systems where conventional order parameters are either insufficient or entirely absent.
\bibliography{ref_ladder}
\end{document}